\documentclass[9pt,conference]{IEEEtran}
\usepackage[flushleft]{threeparttable}

\usepackage{waspaa25} 

\usepackage{bm} 

\usepackage{float}

\usepackage{xcolor}

\title{Learning Perceptually Relevant Temporal Envelope Morphing}

\author{
Carnegie Mellon University \\
\{satvikdixit,laurieheller,chrisdonahue\}@cmu.edu
}

\name{Satvik Dixit,
      Sungjoon Park,
      Chris Donahue$^{\dagger}$,
      Laurie M. Heller$^{\dagger}$\thanks{$^{\dagger}$Co-senior authors.}}
\address{Carnegie Mellon University}

\begin{document}

\maketitle

\begin{abstract}
Temporal envelope morphing, the process of interpolating between the amplitude dynamics of two audio signals, is an emerging problem in generative audio systems that lacks sufficient perceptual grounding. Morphing of temporal envelopes in a perceptually intuitive manner should enable new methods for sound blending in creative media and for probing perceptual organization in psychoacoustics. However, existing audio morphing techniques often fail to produce intermediate temporal envelopes when input sounds have distinct temporal structures; many morphers effectively overlay both temporal structures, leading to perceptually unnatural results. 
In this paper, we introduce a novel workflow for learning envelope morphing with perceptual guidance: we first derive perceptually grounded morphing principles through human listening studies, then synthesize large-scale datasets encoding these principles, and finally train machine learning models to create perceptually intermediate morphs.  
Specifically, we present: (1)~perceptual principles that guide envelope morphing, derived from our listening studies, (2)~a supervised framework to learn these principles, (3)~an autoencoder that learns to compress temporal envelope structures into latent representations, and (4)~benchmarks for evaluating audio envelope morphs, using both synthetic and naturalistic data, and show that our approach outperforms existing methods in producing temporally intermediate morphs.
All code, models, and datasets will be made publicly available upon publication.
\end{abstract}


\section{Introduction}
\label{sec:intro}

\begin{figure*}[htbp]
    \centering
    \includegraphics[width=\linewidth]{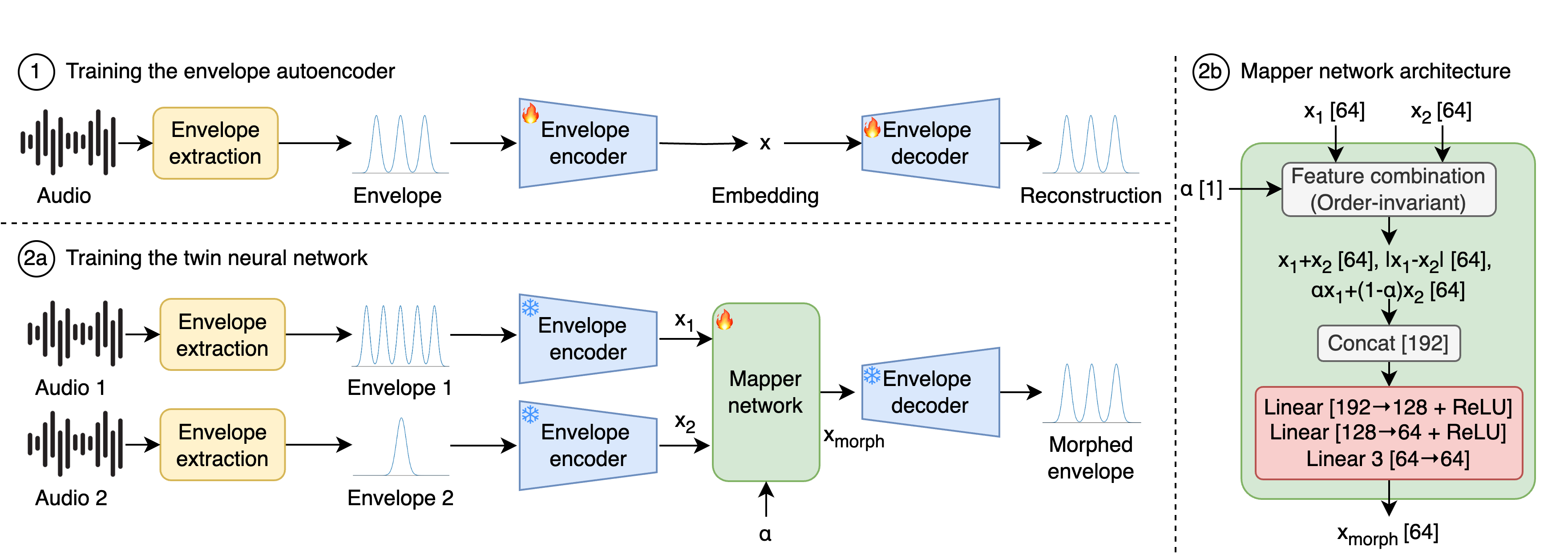} 
    \vspace{-15pt}
    \caption{\textbf{Method overview.}
    In Stage 1, we repurpose an audio autoencoder to learn a latent representation of temporal envelopes. 
    In Stage 2, we train a twin (order-invariant) mapper network (2b) in a supervised setup on synthetic training data that encodes perceptually-grounded morphing rules (2a). During training, the autoencoder is kept frozen while the mapper network learns to morph the envelope embeddings.}
    \label{fig:main} 
    \vspace{-5pt}
\end{figure*}

Kamath et al. define \emph{sound morphing} as ``the process of gradually and smoothly transforming one sound into another to generate novel and perceptually hybrid sounds that simultaneously resemble both''~\cite{kamath2025morphfader}. The ability to smoothly morph between sounds is a longstanding goal in auditory signal processing, with applications ranging from artistic sound design to cognitive science. Such techniques find applications in blending vocal timbres in music production \cite{roddy2014method}, generating transitional sound effects for virtual environments \cite{niu2024soundmorpher} and auditory research \cite{oszczapinska2023underlying}.

Traditional digital signal processing based systems for morphing are limited to musical sounds \cite{donahue2016extended, kazazis2016sound, caetano2019morphing} or vocal sounds \cite{slaney1996automatic, ezzat2005morphing}. These methods employ signal processing techniques to extract features—such as coefficients from a source-filter model representation \cite{slaney1996automatic} or the harmonic components of sounds \cite{kazazis2016sound}—which are then interpolated to produce the morphed outputs. Although these traditional methods perform well for pitched instruments and voiced utterances, such techniques often break down when applied to everyday inharmonic sounds characterized by complex, non-stationary amplitude patterns.

More recent ML-based approaches are based on text-to-audio (TTA) models \cite{liu2023audioldm, liu2024audioldm, kreuk2022audiogen, vyas2023audiobox, huang2023make, ghosal2023text, hung2024tangoflux, hai2024ezaudio, evans2024fast, evans2025stable}. Methods like MorphFader \cite{kamath2025morphfader} and SoundMorpher \cite{niu2024soundmorpher} propose morphing by interpolating between the latent space representations of the two sounds. However, these systems often rely on naive arithmetic averaging in latent space that fails to produce perceptually meaningful results when morphing between sounds with distinct temporal structures — for example, merging a dog's short, impulsive bark with a violin's sustained fluctuations produced by vibrato \cite{niu2024soundmorpher}. For such sounds, the results are often the same as overlaying the two sounds on top of each other.

This failure highlights a gap between  ML-based approaches involving latent space interpolation  and perceptually grounded morphing approaches. At the heart of this challenge lies the \emph{temporal envelope}, which characterizes variations in the overall loudness of a sound over time and constitutes critical information for both identification and perceptual grouping~\cite{moore1997introduction}. Psychoacoustic studies have shown that listeners are sensitive to fine-grained properties of temporal envelopes, such as attack time, impulse spacing, and amplitude modulation \cite{moore1997introduction, McAdams1999}. Yet existing ML systems do not explicitly encode these principles, and accordingly, sound morphing systems based on ML may not handle them in a perceptually intuitive fashion.

In this work, we begin by investigating how humans perceive morphs between temporal envelopes. Through a controlled listening study, we identify key perceptual principles that govern listeners’ sense of continuity and naturalness during audio envelope morphing. These principles are based on simple envelope properties such the quantity of impulses, spacing between impulses, and temporal placement. Building on these insights, we construct a synthetic dataset of envelope tuples—two base envelopes, a perceptually valid ground-truth morph, and an interpolation weight  $\alpha$—by systematically varying these properties one at a time. To encode these envelopes into a compact and perceptually meaningful representation, we train an autoencoder \cite{rumelhart1985learning} that maps raw envelope signals into a low-dimensional latent space. Next, we train a 
``twin'' (order-invariant) 
neural network \cite{bromley1993signature} on our synthetic data to predict the appropriate morph given the two latent representations and $\alpha$. Finally, we create three benchmarks with synthetic and naturalistic envelopes and evaluate our system, demonstrating that our method consistently outperforms existing techniques in producing perceptually intermediate temporal envelope morphs. Our contributions are as follows. 

\begin{itemize}
    \item We derive perceptual principles for envelope morphing from listening studies.
    \item We encode these principles into synthetic data and propose a supervised morphing framework based on a twin neural network.
    \item We train an autoencoder to compress temporal envelope structures into latent representations on envelopes derived from a large-scale dataset of everyday sounds.
    \item We create benchmarks for evaluating audio envelope morphs on synthetic and naturalistic data and demonstrate that our approach is superior to existing methods.
\end{itemize}

\section{Related Work}

\noindent \textbf{Traditional signal processing approaches.}
Early research in sound morphing primarily focused on musical and vocal signals, leveraging parametric signal models to interpolate between sources. Methods based on 
convolution~\cite{dolson1985recent,donahue2016extended}, sinusoidal modeling~\cite{serra1990spectral}, and source-filter decomposition~\cite{slaney1996automatic, rochesso} enabled morphing between pitched sounds by interpolating spectral envelopes, harmonic content, and excitation signals. Extensions to these methods introduced high-level controls over perceptual aspects such as pitch, timbre, and brightness \cite{caetano2011sound}, allowing for greater flexibility in musical applications. However, most of these systems assume stable harmonic structures and voiced excitation, limiting their applicability to the highly diverse and often aperiodic sounds found in everyday environments. 


\noindent \textbf{Psychoacoustic Foundations.}
From a psychoacoustic perspective, temporal envelopes are known to carry essential information for sound identity and perceptual grouping. Moore \cite{moore1997introduction} emphasized that amplitude modulation cues are critical for distinguishing between different sound sources. McAdams \cite{McAdams1999} demonstrated that listeners rely heavily on attack and decay characteristics for recognizing instrument timbres, while Bregman \cite{bregman1994auditory} showed the role of temporal cues in stream formation and segregation. 
McDermott \cite{mcdermott2011sound} showed that statistical properties of temporal envelopes are crucial for identifying complex natural sound textures. More recent work has shown that altering temporal envelopes alone can shift perceptual categorization \cite{oszczapinska2023underlying}, and that both humans and ML models can use temporal and spectral features to identify sound events \cite{dixit2024vision}.
Despite these insights, existing sound event morphing systems have largely overlooked temporal envelopes.

\noindent \textbf{Deep Learning based morphing systems.}
 Recent work has explored latent space interpolation using deep generative models. Niu \cite{niu2024soundmorpher} introduced SoundMorpher, which interpolates in the latent spaces of text-to-audio diffusion models to achieve general-purpose audio morphing. Similarly, Kamath \cite{kamath2025morphfader} proposed MorphFader for interpolating the components of the cross-attention layers to blend audio samples guided by textual prompts. While these methods demonstrate broad domain coverage, they do not provide intuitive morphs of sounds with dissimilar temporal envelopes, such as many rapid impulses versus a few infrequent bursts. This highlights a broader limitation: existing ML-based morphing methods typically fail to work on sounds with distinct temporal dynamics. To address this gap, we develop a temporal envelope morphing framework guided by perceptual principles.

\section{
Perceptually-grounded envelope morphing
}
To establish a perceptually grounded approach to envelope morphing, we conducted listening experiments to identify key principles governing the perceived naturalness of temporal transitions between sounds. Our investigation focused on contrasting a temporally intermediate morph at the \textbf{Midpoint} of two input sounds with two alternative blending strategies: (1) a \textbf{Sequence} created by summing non-overlapping inputs (audio mixing) and (2) an \textbf{Unbalanced} hybrid \cite{caetano2012formal} where one sound's temporal characteristics are applied to another's spectral content, leading to a temporal morph that lacks temporal features of one of the inputs (e.g., chimaeras \cite{smith2002chimaeric, heller2022hybrid}).

We hypothesized that a perceptually intuitive morph would exhibit temporal properties intermediate to those of the two source sounds. We identified three fundamental temporal \emph{dimensions} along which intermediacy could be defined: (1) \textbf{quantity}: the number of salient envelope peaks or discrete auditory events; (2) \textbf{spacing}: the average distance between these events reflecting the perceived pace or density represented by the inter-onset interval (IOI); and (3) \textbf{placement}: the overall temporal positioning of the sound events within a given duration represented by the onset time.

\noindent\textbf{Experimental setup.} Twenty naive listeners were recruited via Prolific.co and participated in the study online, following [institution's] IRB-approved protocols. Participants saw visual and verbal instructions, including examples of morphing with familiar visual stimuli (faces and shapes) \cite{caetano2012formal} to establish an understanding of a morph as a blend exhibiting features of both inputs.

\noindent\textbf{Stimuli.} The auditory stimuli consisted of sequences of 150-ms, 440-Hz pure tones added to a 6-second sample of low-level Gaussian noise. Across trials, we systematically varied one of the three temporal properties described above while holding the others constant. The number of tones (events) ranged from 4 to 16, the inter-onset interval (IOI) varied between 300 and 750 ms, and the onset of the first tone was varied from 10 ms to 3.6 s. Each trial presented two distinct input sounds sequentially, followed by a proposed morphed sound. The Midpoint morph had the linear average of the relevant temporal parameter. The hybrid morphs were temporally intermediate between A and B but were very similar to one of the inputs. The sequential mixdowns were created by sequentially placing the first and second inputs in one audio file. 
Participants rated appropriateness on a 6-point Likert scale (0 = extremely inappropriate, 5 = extremely appropriate).

\noindent\textbf{Results.} Table \ref{perceptual_1} shows the mean appropriateness ratings (and standard deviations) for a selection of trials illustrating each of the three temporal principles and comparing Midpoint morphs to Unbalanced and Sequence alternatives 
In all these conditions, the Midpoint morphs received higher ratings compared to both the Unbalanced and Sequence approaches. (Significant differences in a paired-sample t-test are indicated by asterisks).

\begin{table}
\centering
\caption{Morph appropriateness ratings for different types of morphs on temporal principles. Asterisks indicate that the rating is significantly lower than that of the Midpoint morph. (*\emph{p} $<$ 0.05; **\emph{p} $<$ 0.01; ***\emph{p} $<$ 0.001) }
\label{perceptual_1}
\begin{tabular}{>{\raggedright\arraybackslash}p{0.12\linewidth}>{\centering\arraybackslash}p{0.12\linewidth}>{\centering\arraybackslash}p{0.12\linewidth}|>{\centering\arraybackslash}p{0.11\linewidth}>{\centering\arraybackslash}p{0.11\linewidth}>{\centering\arraybackslash}p{0.11\linewidth}}
\toprule
Principle & Input A & Input B & Midpoint & Unbalanced  & Sequence  \\
\midrule
Quantity\textsuperscript{a}
& 4 events & 8 events& 4.30 (0.92) & 3.40* (1.50) & 2.75*** (1.59) \\[12pt]
Spacing\textsuperscript{b} & 750ms & 350ms & 4.20 (0.83)& 3.30* (1.30)&3.80 (1.44) \\[12pt]
Placement\textsuperscript{c} & 0s & 3.6s & 4.35 (1.04)& 3.40* (1.39)& 3.45** (1.54) \\
\bottomrule
\end{tabular}
\begin{tablenotes}
\item  Properties of Midpoint (M), Unbalanced (U) and Sequence (S) morphs: 
   \item \textsuperscript{a} ($\#$ events): M: 6; U: 5; S: 8
   \item \textsuperscript{b} (IOI): M: 550ms; U: 700ms; S:  750ms then 450ms. 
   \item \textsuperscript{c} (Onset time): M: 1.8s; U: 3.23s, S: 0s. 



  \end{tablenotes}
\end{table}
\vspace{-2pt}
\section{Method} 
\vspace{-2pt}
\begin{figure*}[htbp]
    \centering
    \includegraphics[width=\linewidth]{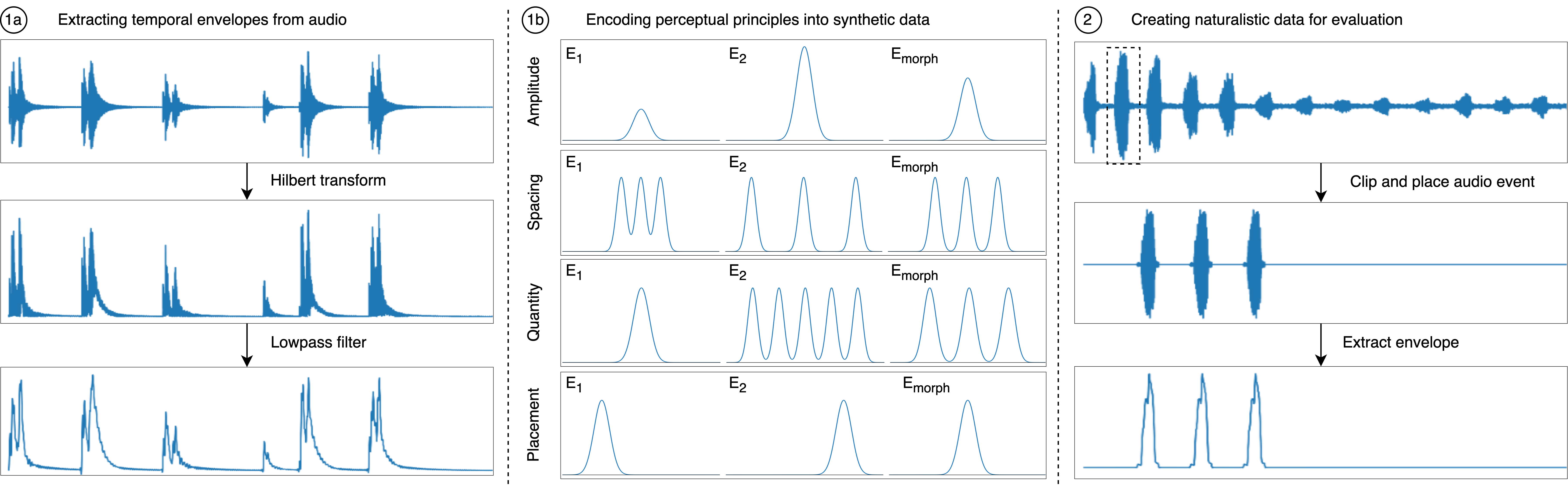} 
    \vspace{-15pt}
    \caption{\textbf{Creating data for unsupervised training, supervised training and evaluation in a naturalistic setting.} We create large scale datasets of (1a) real world audio envelopes for training the autoencoder (1b) synthetic gaussian impulse based envelopes for training the twin neural network on perception based morphing 
    rules and (2) naturalistic audio envelopes to evaluate the envelope morphing systems on real world like sounds.}
    \vspace{-5pt}
    \label{fig:data}
\end{figure*}


Following the results of our perceptual study, here we describe our proposed method to learn a temporal envelope morphing strategy capable of producing intermediate morphs along perceptually-relevant temporal parameters.

\subsection{Unsupervised representation learning for envelopes} \label{subsec:unsupervised}

To morph envelopes, we 
first train an
autoencoder \cite{rumelhart1985learning} to 
learn compact envelop representations, 
providing a bridge to standard morphing methods in deep learning based on interpolation in latent space. 
We extract amplitude envelopes from the AudioCaps dataset \cite{kim2019audiocaps}, a large corpus of 50k $10$s audio clips covering diverse everyday sounds. To extract the temporal envelopes from the audio files, we compute the root mean square (RMS) envelope using the Hilbert transform \cite{he2016praat}. 
This is followed by lowpass filtering with a cutoff frequency of 30 Hz to remove rapid fluctuations and preserve perceptually salient envelope dynamics as shown in Figure \ref{fig:data}. 
This process yields a large-scale dataset of natural audio envelopes.

We repurpose the Stable-Audio open autoencoder architecture \cite{evans2025stable}, consisting of a 5-layer convolutional network, with a 64-dimensional bottleneck latent space. The autoencoder downsamples by a factor of $2048$ in the time dimension. 
We downsample $10$s clips into $2048$ timesteps, resulting in a fixed envelope frame rate of $204.8$Hz. then, the downsampling factor of the architecture results in a single timestep which is projected to an embedding in $\mathbb{R}^{64}$. The autoencoder is trained on the natural audio envelope dataset for 100k steps with a batch size of 64 using a weighted combination of L1 reconstruction loss and an adversarial loss over a discriminator as shown in Figure \ref{fig:main}.

\subsection{Supervised learning for perceptually grounded morphing} \label{subsec:supervised}

While the autoencoder (described in Section \ref{subsec:unsupervised}) learns compact latent representations, we observe that simply extracting the embeddings of two audio envelopes and interpolating between them results in perceptually unintuitive morphs, similar to those obtained with mixdowns. To address the limitations of naive latent interpolation, we propose a supervised framework that learns to predict perceptually appropriate envelope morphs given a pair of input envelopes and an interpolation weight $\alpha$.

We generate a 100k-sample synthetic dataset of envelope morphing tuples designed to capture key perceptual phenomena uncovered in our human studies. 
We formalize the generation of our synthetic morphing dataset by defining a set of functions, 
${\texttt{Dimension} : [0, 1] \to \mathbb{R}^{2048}}$ that map normalized scalar perceptual parameters to synthetic envelopes. 
For example, for the spacing dimension, $\texttt{Spacing}(\alpha)$ synthesizes an envelope of impulses with inter onset interval proportional to $\alpha$. 
Then, we create tuples ${(E_a, E_b, E_{\text{morph}}, \alpha)}$ where 
$E_a = \texttt{Dimension}(0)$, 
$E_b = \texttt{Dimension}(1)$, 
and
$E_{\text{morph}} = \texttt{Dimension}(\alpha)$.
To ensure any network we train using this data focuses solely on learning the morphing rules without being influenced by complex impulse shapes, we use simple Gaussian impulses of width 0.2s as the building blocks for our synthetic envelopes. Each envelope is defined by four perceptual attributes (1)~\textbf{quantity} of impulses, (2)~temporal \textbf{placement}, (3)~\textbf{spacing} between impulses, and (4)~peak \textbf{amplitude} as shown in Figure \ref{fig:data}. We add amplitude as an attribute as it is known to be perceptually relevant~\cite{stevens1957psychophysical}.

Morphs are constructed by systematically varying one parameter at a time in each set, while keeping others fixed. This procedure generates morphing sequences for impulse count changes, temporal position shifts, density changes and amplitude variations. Invalid configurations (e.g., overlapping impulses) are filtered using automated checks. This synthetic dataset enables supervised training with ground-truth morphs based on perceptual information. Our goal is to build an envelope morphing method that can yield intermediate envelopes controlled by $\alpha$, i.e.,~${\texttt{Morph}: E_a, E_b, \alpha \mapsto \hat{E}_{\text{morph}}}$. It is likely that human perception is non-linear with respect to $\alpha$. Perceptual mappings of the form ${\texttt{Map}: [0, 1] \to [0, 1]}$ derived from listening can be incorporated into our method posthoc via ${\texttt{Morph}(E_a, E_b, \texttt{Map}(\alpha))}$.

Our morphing model is a twin neural network \cite{bromley1993signature} trained to predict a morphed embedding from a pair of input envelopes and the interpolation weight $\alpha$. The architecture, illustrated in Figure \ref{fig:main}, consists of a shared pretrained autoencoder encoder (details in Section \ref{subsec:unsupervised}) that maps each envelope to a latent embedding in $\mathbb{R}^{64}$. These two embeddings are then combined using an order-invariant scheme: we compute their sum, their absolute difference, and a weighted average according to $\alpha$. We concatenate these representations to form a feature vector which is passed through a simple three-layer MLP network with a hidden dimension 128. The output is a morphed embedding in $\mathbb{R}^{64}$ which is decoded via the pretrained autoencoder decoder into an output envelope. The model is trained with RMSE loss between the predicted and the ground-truth morph embeddings for 10 epochs with batch size 64.
\vspace{-2pt}
\section{Experiments} 
To assess the effectiveness and generalizability of our proposed audio envelope morphing method, we construct three evaluation benchmarks of increasing complexity. Each benchmark tests a different aspect of perceptual morphing: accuracy on simple synthetic rules, robustness to compositional morphs involving multiple perceptual dimensions, and performance on real-world like acoustic data. In all settings, the task is to predict the morphed envelope given two source envelopes and an interpolation weight $\alpha$, and compare it to the ground truth morph $E_{\text{morph}}$ using RMSE. We evaluate against three baselines: (1) \textbf{audio mixing}, which linearly interpolates between $E_1$ and $E_2$ in amplitude; (2) \textbf{embed mixing}, which interpolates between their latent representations and decodes the result; and (3) \textbf{DTW}, which performs morphing as proposed by \cite{li2023morphing} where Dynamic Time Warping algorithm \cite{muller2007dtw} is used to find the optimal warping path between the two envelopes and linear interpolation is used to generate the morph. 

\subsection{Evaluating morphing along individual perceptual axes}

We construct a benchmark that isolates individual perceptual transformation rules 
using simple gaussian impulse envelopes to evaluate how well the model captures interpretable, low-level morphing behavior. We use 10k synthetic tuples sampled from the same distribution as training (described in Section \ref{subsec:supervised}) but held out from training, with exactly one transformation rule varied in each set. 

\noindent\textbf{Results.} Table \ref{table:syn} shows RMSE across each rule. Audio Mixing achieves the lowest RMSE for amplitude morphing, serving as an oracle in this task where a linear mix directly corresponds to the morph. Our method achieves the lowest overall RMSE, demonstrating particularly strong results on temporal Placement and Quantity.

\begin{table}
\centering
\caption{Evaluating audio envelope morphing methods on synthetic data}
\label{table:syn}
\begin{tabular}{>{\raggedright\arraybackslash}p{0.2\linewidth}>{\centering\arraybackslash}p{0.1\linewidth}>{\centering\arraybackslash}p{0.1\linewidth}>{\centering\arraybackslash}p{0.1\linewidth}>{\centering\arraybackslash}p{0.1\linewidth}>{\centering\arraybackslash}p{0.1\linewidth}}
\toprule
Method & Amplitude & Placement & Spacing & Quantity & All \\
\midrule
Audio Mixing& \textbf{0.003}& 0.105& 0.048& 0.103& 0.063\\
DTW& 0.021& 0.104& 0.047& 0.105& 0.068\\
Embed Mixing& 0.032& 0.099& \textbf{0.046}& 0.102& 0.069\\
Ours& 0.036& \textbf{0.077}& 0.050& \textbf{0.086}& \textbf{0.062}\\ 
\bottomrule
\end{tabular}
\end{table}

\subsection{Evaluating the generalization to compositional morphing}

To test whether the model (trained on individual perceptual rules in isolation) can generalize to more complex morphs that simultaneously involve multiple perceptual axes (e.g., changing both density and amplitude), we construct a second benchmark consisting of 10k synthetic triplets. Each test sample involves changing 2, 3 or all 4 rule dimensions. These configurations are held out from training to simulate compositional generalization and assess whether the model’s internal representation captures latent structure across rules. 

\noindent\textbf{Results.} Table~\ref{table:composed} reports the RMSE for various combinations of simultaneously varying perceptual rules. Notably, when amplitude is composed with other features, audio mixing is no longer an oracle and our proposed method consistently outperforms all baselines across all compositional morphing scenarios. The overall performance of our method on these complex, held-out compositional morphs is on par with its performance on the isolated perceptual axes, demonstrating strong evidence of generalization.

\begin{table}
\centering
\caption{Evaluating envelope morphing methods on composed synthetic data}
\label{table:composed}
\begin{tabular}{>{\centering\arraybackslash}p{0.09\linewidth}>{\centering\arraybackslash}p{0.09\linewidth}>{\centering\arraybackslash}p{0.09\linewidth}>{\centering\arraybackslash}p{0.09\linewidth}>{\raggedright\arraybackslash}p{0.06\linewidth}>{\raggedright\arraybackslash}p{0.06\linewidth}>{\raggedright\arraybackslash}p{0.06\linewidth}>{\centering\arraybackslash}p{0.06\linewidth}}
\toprule
Amplitude & Placement & Spacing & Quantity & Audio Mix & DTW & Embed Mix & Ours \\
\midrule
\checkmark & \checkmark & & & 0.103& 0.102& 0.098& \textbf{0.077}\\
\checkmark & & \checkmark & & 0.049& \textbf{0.048}& \textbf{0.048}& 0.052\\
\checkmark & & & \checkmark & 0.114& 0.116& 0.113& \textbf{0.094}\\
& \checkmark & \checkmark & & 0.107& 0.106& 0.102& \textbf{0.080}\\
& \checkmark & & \checkmark & 0.120& 0.119& 0.114& \textbf{0.093}\\
& & \checkmark & \checkmark & 0.117& 0.116& 0.111& \textbf{0.094}\\
\midrule
\checkmark & \checkmark & \checkmark & & 0.110& 0.109& 0.105& \textbf{0.085}\\
& \checkmark & \checkmark & \checkmark & 0.127& 0.125& 0.123& \textbf{0.101}\\
\checkmark & & \checkmark & \checkmark & 0.123& 0.121& 0.117& \textbf{0.101}\\
\checkmark & \checkmark & & \checkmark & 0.122& 0.120& 0.118& \textbf{0.099}\\
\midrule
\checkmark & \checkmark & \checkmark & \checkmark & 0.121& 0.120& 0.118& \textbf{0.098}\\
\bottomrule
\end{tabular}
\end{table}

\vspace{-2pt}
\subsection{Evaluating the generalization to naturalistic signals}
\vspace{-2pt}

While synthetic data offers interpretability and controlled supervision, it lacks the acoustic complexity and variability of real-world audio. However, fully natural morphing benchmarks lack ground-truth morphs. To balance these needs, we design a naturalistic audio envelope benchmark using real world impulse sounds arranged with synthetic structure \cite{xie2024audiotime} as shown in Figure \ref{fig:data}. We extract diverse impulse events (e.g., a dog bark, a bicycle bell etc.) from the Clotho dataset \cite{drossos2020clotho}, manually isolate single impulses, and use them to generate a set of 10k tuples following the same perceptual morphing rules as described in Section~\ref{subsec:supervised}). These audio clips are processed into envelope representations using the same pipeline as before.  

\noindent\textbf{Results.} Table~\ref{table:quant_hybrid} shows that similar to the synthetic individual axis evaluation, Audio Mixing performs well on amplitude morphing. However, our proposed method demonstrates robust generalization to real-world sound envelopes, achieving the lowest overall RMSE, highlighting its potential for creating perceptually plausible envelope morphs for a wide range of real-world audio.

\begin{table}
\centering
\caption{Evaluating audio envelope morphing methods on naturalistic data}
\label{table:quant_hybrid}
\begin{tabular}{>{\raggedright\arraybackslash}p{0.2\linewidth}>{\centering\arraybackslash}p{0.1\linewidth}>{\centering\arraybackslash}p{0.1\linewidth}>{\centering\arraybackslash}p{0.08\linewidth}>{\centering\arraybackslash}p{0.08\linewidth}>{\centering\arraybackslash}p{0.08\linewidth}}
\toprule
Method & Amplitude & Placement & Spacing & Quantity & All \\
\midrule
Audio Mixing& \textbf{0.002}& 0.074&  0.045& 0.084& 0.052\\
DTW& 0.024& 0.073& 0.044& 0.086& 0.057\\
Embed Mixing& 0.029& 0.068& \textbf{0.040}&  0.084& 0.056\\
Ours& 0.032&\textbf{ 0.058}& 0.043& \textbf{0.071}& \textbf{0.051}\\ 
\bottomrule
\end{tabular}
\end{table}

\vspace{-4pt}
\section{Conclusion}
\vspace{-2pt}
    
We introduced a perceptually grounded framework for morphing temporal envelopes, guided by empirical studies on human auditory perception. By codifying key perceptual factors—such as the quantity, spacing, and placement of impulses—we built a large synthetic dataset of envelope morphing examples to train a twin network for perceptually valid interpolations. Experiments show our framework consistently outperforms conventional methods. As modern synthesis and generative models increasingly rely on temporal envelopes \cite{chung2024t, garcia2025sketch2sound, gramaccioni2024stable, lee2024video}, our approach offers a perceptually intuitive way to morph them, enhancing controllability and naturalness. Bridging psychoacoustics and deep learning, this work advances audio generation toward perceptually grounded morphing with applications in assistive sound design, immersive media, and computational hearing.


\section{Acknowledgment}
\label{sec:ack}

We thank Sony for their support of this work through the Sony Research Award Program (RAP).


\bibliographystyle{IEEEtran}
\bibliography{refs25}







\end{document}